\documentclass[preprint2]{aastex}

\usepackage{color}

\shorttitle{MEASUREMENTS OF DIFFUSE SKY EMISSIONS AT $3.5$ and $4.9\,\rm{\mu m}$}

\shortauthors{Sano et al.}

\begin{document}


\title{MEASUREMENTS OF DIFFUSE SKY EMISSION COMPONENTS IN HIGH GALACTIC LATITUDES AT $3.5$ and $4.9\,\rm{\mu m}$ USING DIRBE AND WISE DATA}


\author{K. SANO\altaffilmark{1,2}, K. KAWARA\altaffilmark{3}, S. MATSUURA\altaffilmark{4}, H. KATAZA\altaffilmark{1,2}, \\
T. ARAI\altaffilmark{5}, AND Y. MATSUOKA\altaffilmark{6}}
\affil{
\altaffilmark{1}Department of Astronomy, Graduate School of Science, The University of Tokyo, \\
Hongo 7-3-1, Bunkyo-ku, Tokyo 113-0033, Japan\\
\altaffilmark{2}Institute of Space and Astronautical Science, Japan Aerospace Exploration Agency,\\
3-1-1 Yoshinodai, Chuo-ku, Sagamihara, Kanagawa 252-5210, Japan\\
\altaffilmark{3}Institute of Astronomy, University of Tokyo, 2-21-1, Osawa, Mitaka, Tokyo 181-0015, Japan \\
\altaffilmark{4}Department of Physics, School of Science and Engineering, Kwansei Gakuin University, 2-1 Gakuen, Sanda, Hyogo 669-1337, Japan \\
\altaffilmark{5}Frontier Research Institute for Interdisciplinary Science, Tohoku University, Sendai 980-8578, Japan \\
\altaffilmark{6}National Astronomical Observatory of Japan, 2-21-1 Osawa, Mitaka, Tokyo 181-8588, Japan \\
}

\email{sano@ir.isas.jaxa.jp}






\begin{abstract}
Using all-sky maps obtained from {\it COBE}/DIRBE at $3.5$ and $4.9\,\rm{\mu m}$, we present a reanalysis of diffuse sky emissions such as zodiacal light (ZL), diffuse Galactic light (DGL), integrated starlight (ISL), and isotropic residual emission including the extragalactic background light (EBL).
Our new analysis, which includes an improved estimate of ISL using the Wide-field Infrared Survey Explorer (WISE) data, enabled us to find the DGL signal in a direct linear correlation between diffuse near-infrared and $100\,\rm{\mu m}$ emission at high Galactic latitudes ($|{\it b}| > 35^\circ$).
At $3.5\,\rm{\mu m}$, the high-latitude DGL result is comparable to the low-latitude value derived from the previous DIRBE analysis.
In comparison with models of the DGL spectrum assuming a size distribution of dust grains composed of amorphous silicate, graphite, and polycyclic aromatic hydrocarbon (PAH), the measured DGL values at $3.5$ and $4.9\,\rm{\mu m}$ constrain the mass fraction of PAH particles in the total dust species to be more than $\sim 2\%$.
This was consistent with the results of {\it Spitzer}/IRAC toward the lower Galactic latitude regions.
The derived residual emission of $8.9\pm3.4\,\rm{nWm^{-2}sr^{-1}}$ at $3.5\,\rm{\mu m}$ is marginally consistent with the level of integrated galaxy light and the EBL constraints from the $\gamma$-ray observations.
The residual emission at $4.9\,\rm{\mu m}$ is not significantly detected due to the large uncertainty in the ZL subtraction, same as previous studies.
Combined with our reanalysis of the DIRBE data at $1.25$ and $2.2\,\rm{\mu m}$, the residual emission in the near-infrared exhibits the Rayleigh-Jeans spectrum.

\end{abstract}





\section{INTRODUCTION}

\subsection{Current Measurements of the Near-Infrared Extragalactic Background Light}

Extragalactic background light (EBL) is the cumulative light emitted by any radiation process between the reionization era and the present epoch.
Sources include nucleosynthesis, emission from heated dust, and exotic particle decay.
In particular, near-infrared (IR) EBL is thought to contain the redshifted ultraviolet radiation, which may have contributed to the reionization of the universe.
Therefore, near-IR EBL can serve as an indicator in investigations of galaxy formation and evolution in the early universe.

In the absolute measurement of near-IR EBL, foreground emissions, such as zodiacal light (ZL), integrated starlight (ISL), and diffuse Galactic light (DGL), must be removed from the sky brightness.
Previously, near-IR EBL has been investigated using data obtained from space telescopes, such as the Diffuse Infrared Background Experiment (DIRBE) aboard the Cosmic Background Explorer ({\it COBE}) satellite (e.g., Hauser et al. 1998; Gorjian et al. 2000; Cambr\'esy et al. 2001; Levenson et al. 2007; Levenson \& Wright 2008), 
the Infrared Telescope in Space ({\it IRTS}) (Matsumoto et al. 2005; Matsumoto et al. 2015), and {\it AKARI} (Tsumura et al. 2013c).
The literature has shown that particularly at 1--$2\, \mu \rm m$, the residual intensity was $\sim 2$--5 times higher than the intensity of the integrated galaxy light (IGL) derived from deep galaxy counts (Madau \& Pozzetti 2000; Totani et al. 2001).
In addition, Sano et al. (2015), hereafter Paper I, reanalyzed the DIRBE data and revealed that deviations of the residual emission from isotropy are less than 10\% at $1.25$ and $2.2\, \mu \rm m$ in the entire sky.

Another indirect estimation of EBL intensity has been derived by observation of the spectra of blazars emitting high-energy $\gamma$-rays.
Since such $\gamma$-rays are attenuated by interacting with EBL photons in the propagation of the intergalactic space, upper limits of EBL intensity can be derived by assuming the intrinsic spectra of blazars
 (e.g., Guy et al. 2000; Dwek \& Krennrich 2005; Schroedter 2005; Aharonian et al. 2006; Albert et al. 2008; Mazin \& Raue 2007; Orr et al. 2011; Meyer et al. 2012). 
In addition, the recent result of the High Energy Stereoscopic System (H.E.S.S.) experiment has offered direct constraints on EBL, rather than estimated upper limits (Abramowski et al. 2013).
Most of the $\gamma$-ray constraints on EBL at 1--$2\, \mu \rm m$ are a few to several times lower than the derived residual emissions, indicating that not all of the excess brightness originates from extragalactic sources.

Several studies have estimated the contributions of exotic extragalactic sources other than IGL to EBL, including population III (Pop-III) stars (e.g., Inoue et al. 2013; Fernandez et al. 2013), intrahalo light (IHL) (Cooray et al. 2012; Zemcov et al. 2014), direct collapse black holes (Yue et al. 2013), and dark stars (Maurer et al. 2012).
As suggested by Paper I, the total contribution of all of these sources cannot account for the observed excesses at $1.25$ and $2.2\, \mu \rm m$.
This fact also supports the idea that a portion of the excess emission may originate from the local universe: the Milky Way and/or the solar system.

At around $3.5\,\rm{\mu m}$, the previously derived residual emissions are smaller than those at 1--$2\, \mu \rm m$, approaching IGL and $\gamma$-ray upper limits (Gorjian et al. 2000; Wright \& Reese 2000; Levenson et al. 2007; Levenson \& Wright 2008). 
Since the ZL contribution is much more dominant in wavelengths longer than $4\,\rm{\mu m}$, residual intensity has only been derived as upper limits (Hauser et al. 1998;  Tsumura et al. 2013c), consistent with the derived IGL level (Fazio et al. 2004).

\subsection{Near-Infrared Diffuse Galactic Light}
 
DGL, one of the foregrounds of optical to near-IR EBL, is thought to consist of starlight scattered by interstellar dust grains and thermal emission from the grains heated by this starlight.
DGL can be used to measure the properties of the interstellar environment, including the intensity of the interstellar radiation field (ISRF),  the size distribution of the dust grains, and the mass ratio of the total grains to the interstellar medium (ISM).
In the current interstellar dust models (e.g., Li \& Draine 2001; Draine \& Li 2007; Compi\`egne et al. 2011), interstellar dust is assumed to be a mixture of silicate and graphitic grains, including polycyclic aromatic hydrocarbon (PAH).

At optical wavelengths, DGL has been quantitatively investigated as a component that linearly correlates with diffuse interstellar $100 \,\rm{\mu m}$ emission (e.g., Matsuoka et al. 2011; Brandt \& Draine 2012; Ienaka et al. 2013).  
In contrast, the DGL in the near-IR has been difficult to measure because the optical depth is too low to enhance the DGL emission (Leinert et al. 1998).
It was only recently that an analysis of the data obtained by the Cosmic Infrared Background Experiment (CIBER) (Arai et al. 2015) and DIRBE (Paper I) found the DGL component in high Galactic latitudes at 1--$2\, \mu \rm m$.

In the previous analysis of the DIRBE data, Arendt et al. (1998) used only a Faint Source Model based on Wainscoat et al.'s (1992)  star-counts model for the ISL evaluation and did not find the DGL components at $1.25$ and $2.2\, \mu \rm m$.
In contrast, Paper I reanalyzed the DIRBE data with an improved ISL evaluation using the Two Micron All-Sky Survey (2MASS) Point Source Catalog (PSC) data (Cutri et al. 2003; Skrutskie et al. 2006) and found the high-latitude DGL component at $1.25$ and $2.2\, \mu \rm m$.
This suggests that precise ISL evaluation is required for the DGL measurement because the brightnesses of  both components are expressed as similar functions of the Galactic latitude in principle.
Combining the results with those in the optical wavelengths (Matsuoka et al. 2011; Ienaka et al. 2013), Arai et al. (2015) and Paper I reported a bluer DGL spectrum from the optical to $\sim 2\, \mu \rm m$ wavelengths, which is close to the expected scattered spectrum of an interstellar dust model in which the smaller dust grains are dominant.
Typical intensities of the high-latitude DGL at these bands were found to be less than 10\% of that of the residual emission (Paper I).

In the longer near-IR wavelengths, Arendt et al. (1998) derived the DGL results at $3.5$ and $4.9\, \mu \rm m$ from the DIRBE data only at low-Galactic ($|{\it b}| < 30^\circ$) and high-ecliptic latitudes ($|{\it \beta}| > 40^\circ$) to enhance the dust emission and avoid strong ZL emission.
Because of its low precision in the ISL evaluation, their results were derived not by direct correlation against diffuse $100 \,\rm{\mu m}$ emission but by the color-color method, which makes use of the color difference between ISL and DGL.
More importantly, they assumed that the DGL results in the low Galactic latitudes, including the region close to the Galactic plane, are the same as those in the high-latitude region of greater interest for DGL and EBL measurements.
Naturally, it is questionable whether the DGL results at low latitudes are applicable at high latitudes because dust properties, such as size distribution and composition of the grains, could be different between these regions.
Thereafter, the obtained DGL results have been used for the high-latitudes DGL contribution to the EBL measurements (Dwek \& Arendt 1998; Gorjian et al. 2000).

Interstellar dust grains heated by the ISRF for the solar neighborhood, which is estimated by Mathis et al. (1983),  are expected to radiate thermal emission at the $3.5$ and $4.9\, \mu \rm m$ wavelength range (e.g., Dwek et al. 1997; Li \& Draine 2001; Draine \& Li 2007; Compi\`egne et al. 2011).
In contrast, Tsumura et al. (2013b) and Matsumoto et al. (2015) recently reported no detection of DGL at high Galactic latitudes ($|{\it b}| > 30^\circ$) in their analysis of the diffuse sky spectrum obtained with {\it AKARI} and {\it IRTS}, because of the low signal-to-noise ratio in their analyses of limited regions of the sky.
If the ISL estimation is improved from the previous DIRBE analysis (Arendt et al. 1998), we expect DIRBE to be the most appropriate data to confirm the high-latitude DGL component since they provide us with all-sky maps of higher signal-to-noise ratio.

\subsection{Motivation of the Present Study}

As described in Paper I, it is beneficial to use a combination of all-sky maps obtained from DIRBE and the deep point source catalog (2MASS PSC) for ISL evaluation, in order to analyze the diffuse sky emission components at $1.25$ and $2.2\, \mu \rm m$.
To apply this strategy to the DIRBE data in the $3.5$ and $4.9\, \mu \rm m$ bands, we use the AllWISE source catalog based on the Wide-field Infrared Survey Explorer mission (WISE; Wright et al. 2010).
The AllWISE source catalog provides accurate photometries and positions of over 747 million objects in the $3.4$ (W1), $4.6$ (W2), $12$ (W3), and $22 \,\rm{\mu m}$ (W4) bandpasses.  
In the W1 and W2 bands, the AllWISE achieved $5\sigma$ sensitivities of $16.9$ and $16.0\,$mag in the Vega magnitude, respectively.
These are comparable to those of the 2MASS PSC --- 15.8 and $14.3\,$mag in the $J$ and $K_s$ bands, respectively.
Since no star-count catalog has been available at $3.5\, \mu \rm m$,  previous studies have used the $2.2\, \mu \rm m$ DIRBE or 2MASS data to estimate the ISL contribution at $3.5\, \mu \rm m$ (Dwek \& Arendt 1998; Wright \& Reese 2000; Levenson et al. 2007).
In contrast, the present study directly derives the ISL intensity at $3.5$ and $4.9\, \mu \rm m$ using the AllWISE source catalog. 

Using the DIRBE data, the present study mainly describes reanalysis of DGL and the residual emission components at $3.5$ and $4.9\, \mu \rm m$ at high Galactic latitudes.
As our results, we first found a direct linear correlation between the diffuse near-IR and interstellar $100\, \mu \rm m$ emission at high latitudes ($|{\it b}| > 35^\circ$), which allows us to compare our results with those predicted by dust models in the high-latitude region.
At $3.5\, \mu \rm m$, the intensity of the residual emission was the same level as in the other studies within the uncertainty, marginally consistent with the IGL level and $\gamma$-ray constraints on EBL.

The remainder of this paper is organized as follows.
In Section 2, we provide an overview of the DIRBE data used in this paper and describe the analysis, including the decomposition of each diffuse emission.
In Section 3, we present the results of each emission component at $3.5$ and $4.9\, \mu \rm m$ in the high-latitude region.
We compare the DGL results with the predictions of interstellar dust models composed of silicate and carbonaceous grains including PAHs.
We also compare the derived residual emission components with other studies.
A summary of this paper appears in Section 4.

In this paper, the intensities of the sky emissions are expressed in units of $\rm{nWm^{-2}sr^{-1}}$ or $\rm{MJy\,sr^{-1}}$.
The conversion formula between these units is
\begin{equation}
\it{\nu I_{\nu}} \,(\rm{nWm^{-2}sr^{-1}}) = [3000/{\it\lambda} \,(\rm{\mu m})]\, {\it I_{\nu}}\, (\rm{MJy\,sr^{-1}}).
\end{equation}





\begin{figure*}
\begin{center}
 \includegraphics[scale=0.8]{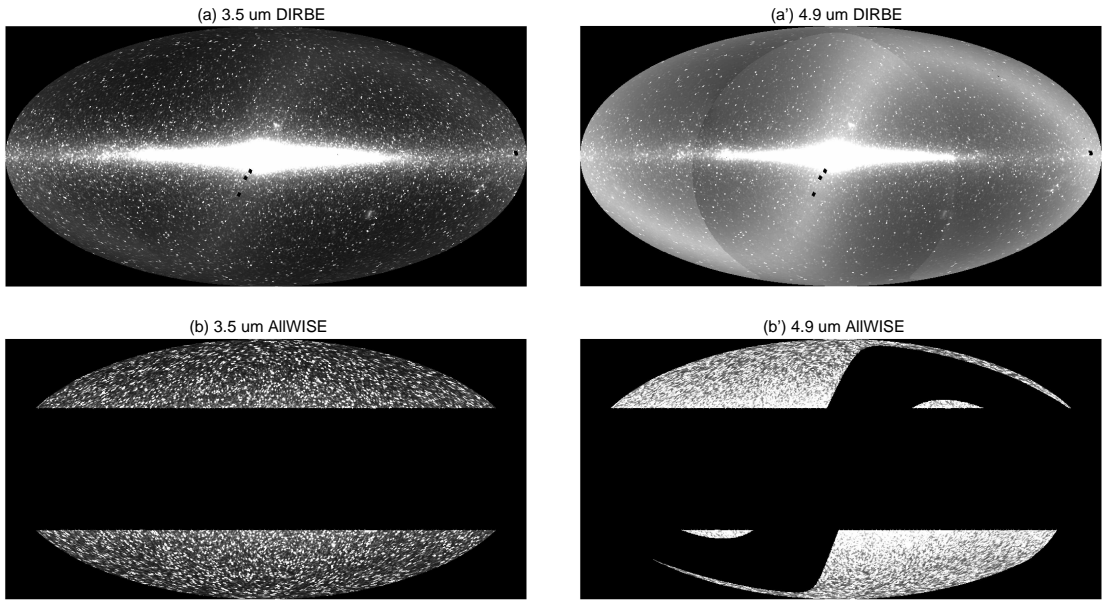} 
 \caption
 {Mollweide projections of the all-sky maps at $3.5$ (left panels) and $4.9\, \mu \rm m$ (right panels) in Galactic coordinates with the Galactic center in the middle.
 Panels (a) and (a') illustrate the all-sky DIRBE ${\it \epsilon} = 90^\circ$ intensity maps.
 Panels (b) and (b') are integrated intensity maps of the AllWISE sources at high Galactic latitudes ($|{\it b}|>35^\circ$), created as described in subsection 2.2.3.
At $4.9\, \mu \rm m$, the region of low ecliptic latitudes ($|{\it \beta}|<20^\circ$) is also excluded in the present analysis because Kelsall et al. (1998) reports the incompleteness of the ZL model in that region.
Each map is arbitrarily scaled for illustration.
}
\end{center}
\end{figure*}

\section{DATA ANALYSIS}

\subsection{DIRBE Data}

DIRBE was one of the instruments onboard the {\it COBE} spacecraft and was designed to investigate the near to far-IR EBL.
Its cryogenic operation was implemented from 1989 November 24 to 1990 September 21.
During these 10 months, almost the entire sky was observed in 10 photometric bands with effective wavelengths of $1.25$, $2.2$, $3.5$, $4.9$, $12$, $25$, $60$, $100$, $160$, and $240\, \mu \rm m$.
The DIRBE instrument was designed to conduct accurate absolute measurements of the sky brightness, with a stray light rejection of less than $1\, \rm{nWm^{-2}sr^{-1}}$ (Magner 1987) and an absolute gain calibration uncertainty of $3.1\%$ and $3.0\%$ at $3.5$ and $4.9\, \mu \rm m$, respectively (Hauser et al. 1998).
Consequently, the all-sky maps at IR wavelengths were created with a beam size of $\sim 0.7^\circ \times 0.7^\circ$.

DIRBE observed every line of sight of the sky with a solar elongation angle $({\it \epsilon})$ of $90^\circ$ once every 6 months; once or twice during the 10-month observation.
From the ${\it \epsilon} = 90^\circ$ maps, the sky brightness at each pixel can be obtained by interpolating the observations made at various times at ${\it \epsilon}$ close to $90^\circ$.
As described in subsection 2.2.1, the present analysis includes the evaluation of the scaling factor of the DIRBE ZL model (Kelsall et al. 1998; hereafter the ``Kelsall model'') against the DIRBE data themselves.
We then use the ${\it \epsilon} = 90^\circ$ maps from which the ZL contribution is not subtracted.
The ${\it \epsilon} = 90^\circ$ maps provide both the sky coordinates and the observation date for each pixel, which are necessary for running the code of the Kelsall model.
In contrast, Zodi-Subtracted Mission Average (ZSMA) maps used in the previous studies (Arendt et al. 1998; Cambr\'esy et al. 2001) provided only the sky coordinates.
The Kelsall model therefore cannot be used in the analysis of the ZSMA maps.
For this reason, we use the ${\it \epsilon} = 90^\circ$ maps created by 6 month observations, starting from 1990 January 1. 

Panels (a) and (a') of Figure 1 depict the ${\it\epsilon} = 90^\circ$ maps at $3.5$ and $4.9\, \mu \rm m$, respectively, on a Mollweide projection that is reprojected from the original ``{\it COBE} Quadrilateralized Spherical Cube'' (CSC) projection adopted in DIRBE products.
The CSC projection is an approximately equal-area projection of the celestial sphere onto an inscribed cube.
Each cube face is divided into $256\times 256$ pixels; thus, all-sky maps have $256^2\times 6 = 393216$ pixels, whose scales are approximately $0.32^\circ \times 0.32^\circ$.
The following analysis is performed on the CSC projection maps.
The ${\it\epsilon} = 90^\circ$ maps and the beam profile maps used in this paper are available at the DIRBE website, ``lambda.gsfc.nasa.gov/product/cobe/''.

\subsection{Model of Diffuse Sky Emissions}

In the following analysis, the intensity of the DIRBE ${\it \epsilon} = 90^\circ$ map, ${\it I_i(\rm{Obs})}$, is modeled as ${\it I_i(\rm{Model})}$, where the subscript ``${\it i}$'' refers to one of the two bands ($3.5$ or $4.9\, \mu \rm m$).
The sky brightness is assumed to be a linear combination of four emission components, i.e., the ZL, DGL, ISL, and residual emissions including EBL.
The ${\it I_i({\rm Model})}$ is therefore described as
\begin{equation}
{\it I_i({\rm Model}) = I_i({\rm ZL}) + I_i({\rm DGL}) + I_i({\rm ISL}) + I_i({\rm Resid})},
\end{equation}
where ${\it I_i(\rm{ZL})}$, ${\it I_i(\rm{DGL})}$, ${\it I_i(\rm{ISL})}$, and ${\it I_i(\rm{Resid})}$ indicate the intensities of the ZL, DGL, ISL, and residual emission, respectively.
These four components are modeled as described below.

\subsubsection{Zodiacal Light}

The ZL term $I_i({\rm ZL})$ is defined as
\begin{equation}
{\it I_i({\rm ZL}) = a_iI_i({\rm Kel})},
\end{equation}
where ${\it a_i}$ is a free parameter and ${\it I_i({\rm Kel})}$ is the ZL intensity estimated by the Kelsall model with a default set of parameters (Kelsall et al. 1998).
The Kelsall model is a parameterized physical model of interplanetary dust emission fitted to the seasonal variation in the sky brightness observed by DIRBE.
To evaluate the typical scaling factor of the Kelsall model against the DIRBE data, we introduce the free parameter ${\it a_i}$.
If the Kelsall model completely reproduces the seasonal variation observed by DIRBE, the parameter ${\it a_i}$ is determined to be 1.0.

\subsubsection{Diffuse Galactic Light}

Diffuse $100\,\rm{\mu m}$ emission from interstellar dust has been suggested as a physically appropriate tracer of DGL (Brandt \& Draine 2012).
In fact, several studies have reported a good linear correlation between the intensities of diffuse optical to near-IR light and $100\,\rm{\mu m}$ emission (e.g., Matsuoka et al. 2011; Ienaka et al. 2013; Arai et al. 2015; Paper I).
In their analysis of the translucent cloud MBM32 in the optical wavelengths, Ienaka et al. (2013) suggested that the linear correlation may break in the optically thick region where DGL itself suffers extinction by interstellar dust.
In contrast, at 1--$2\,\rm{\mu m}$, Arai et al. (2015) and Paper I reported a linear correlation in the wide range of the $100\,\rm{\mu m}$ emission intensity, $\sim0$--$10\,{\rm MJy\,sr^{-1}}$, because optically thin fields are dominant in the high Galactic latitudes in the near-IR bands.
We therefore expect the diffuse sky brightness at $3.5$ and $4.9\,\rm{\mu m}$ to show linear correlation against diffuse $100\,\rm{\mu m}$ emission.

In the present analysis, we use the diffuse $100\,\rm{\mu m}$ emission map created by Schlegel et al. (1998), hereafter referred to as the ``SFD map''. 
The SFD map was created using the data collected with the Infrared Astronomical Satellite ({\it IRAS}) and DIRBE, from which point sources and the ZL contribution were removed.
The different pixel scales between the SFD map ($\sim 0.04^\circ \times 0.04^\circ$) and the ${\it\epsilon} = 90^\circ$ DIRBE maps ($\sim 0.32^\circ \times 0.32^\circ$) cause photometric bias in the correlation analysis.
We therefore apply an $8 \times 8$ pixel binning to the SFD map to obtain the same spatial resolution as that of the DIRBE ${\it\epsilon} = 90^\circ$ maps.
The DGL term ${\it I_i({\rm DGL})}$ is then defined as
\begin{equation}
{\it I_i({\rm DGL}) = b_iI_{{\rm 100}}},
\end{equation}
where ${\it b_i}$ is a free parameter and $I_{100}$ is the interstellar $100\,\rm{\mu m}$ emission intensity, defined as
\begin{equation}
I_{100} = I_{{\rm SFD}}-0.78\, \rm{MJy\,sr^{-1}},
\end{equation}
with $I_{{\rm SFD}}$ being the diffuse $100\,\rm{\mu m}$ emission intensity of the binned SFD map.
Lagache et al. (2000) derived the $100\,\rm{\mu m}$ EBL intensity of $0.78\,\pm\,0.21\,\rm{MJy\,sr^{-1}}$.
In addition, Matsuoka et al. (2011) showed the intensity relation of the SFD map and the diffuse optical light observed by {\it Pioneer 10/11}, and found a correlation break below $\sim 0.8\,\rm{MJy\,sr^{-1}}$ on the SFD map.
Based on these results, we assume the isotropic EBL at $100\,\rm{\mu m}$ to be $0.78\,\rm{MJy\,sr^{-1}}$, and subtract this from the intensity of the SFD map to obtain the $100\,\rm{\mu m}$ intensity associated with interstellar dust.

\subsubsection{Integrated Starlight}

To estimate the ISL contribution in the ${\it\epsilon} = 90^\circ$ maps, we use the AllWISE source catalog created by the WISE mission (Wright et al. 2010) because the W1 ($3.4 \,\rm{\mu m}$) and W2 ($4.6 \,\rm{\mu m}$) bands are close to the DIRBE bands that interest us.
The AllWISE catalog contains point and extended sources with the $5\sigma$ sensitivities of $16.9$ and $16.0\,$mag in the Vega magnitude in the W1 and W2 bands, respectively.

To calculate the ISL intensity at each pixel in the DIRBE maps, we need to know beam profiles of the ${\it \epsilon} = 90^\circ$ maps.
The intensity of the DIRBE ${\it \epsilon} = 90^\circ$ map is derived as the average of dozens of observations.
Therefore, we should use an averaged beam of the daily map created in the manner described in Section 3.1.3 and Figure 2 of Paper I.
The shape of the averaged beam at $3.5$ and $4.9\, \mu \rm m$ is similar to that at $1.25$ and $2.2\, \mu \rm m$, with a full width at half maximum of $\sim 1^\circ$.

Because the spectral response function of WISE is different from that of DIRBE, we need to estimate the flux densities of the sources in the DIRBE bands at $3.5$ and $4.9\, \mu \rm m$ from those in the W1 and W2 bands, respectively. 
In the wavelength range that interests us, a vast majority of the Galactic sources exhibit the Rayleigh-Jeans spectrum;
\begin{equation}
F_{\nu} \propto \nu^2,
\end{equation}
where $F_{\nu}$ is the flux density per unit frequency.
Then, the conversion formula between the weighted-mean flux density in the WISE band ($F^{W_i}$) and that in the DIRBE band ($F^{D_i}$) is described as
\begin{equation}
F^{D_i} = \biggl(\frac{\int F_{\nu} R^{D_i}_{\nu} /\nu \, d\nu}{\int R^{D_i}_{\nu} /\nu \, d\nu} \bigg/ \frac{\int F_{\nu} R^{W_i}_{\nu}  /\nu \, d\nu}{\int R^{W_i}_{\nu} /\nu \, d\nu}\biggr) F^{W_i}
\end{equation}
\begin{equation}
= \alpha_i F^{W_i},
\end{equation}
where $R^{D_i}_{\nu}$ and $R^{W_i}_{\nu}$ are the spectral response functions of DIRBE and WISE in the $i$ band, respectively, taken from the COBE/DIRBE explanatory supplement (1998) and Wright et al. (2010).
Equation (7) adopts the formula of flux density at the isophotal wavelength, which was defined in Tokunaga \& Vacca (2005). 
The derived conversion terms, $\alpha_i$, are 0.902 and 0.882 at $3.5$ and $4.9\, \mu \rm m$, respectively.
As described in Wright et al. (2010), $F^{W_i}$ is defined as
\begin{equation}
F^{W_i} = F^{W_i}_{0} 10^{-0.4m_i},
\end{equation}
where $m_i$ is the magnitude of the source in the AllWISE catalog and $F^{W_i}_{0}$ is the zero magnitude at the WISE photometric system --- 306.681 and 170.663 Jy in the W1 and W2 bands, respectively.
The magnitude of the AllWISE source is derived under an assumed source spectrum of $F_{\nu} \propto \nu^{-2}$ (Wright et al. 2010).
As described in Table 1 of Wright et al. (2010), the difference of the flux density between the sources of the Rayleigh-Jeans spectrum $F_{\nu} \propto \nu^2$ and $F_{\nu} \propto \nu^{-2}$ is less than 1\%.
Similarly, the intensity of the DIRBE map is estimated by assuming $F_{\nu} \propto \nu^{-1}$ (COBE/DIRBE explanatory supplement 1998), and the divergence from the spectrum of $F_{\nu} \propto \nu^2$ is less than $\sim 2\%$.
These differences associated with the color correction are small in comparison to the ISL term derived in the following section.

In the AllWISE catalog, we should select only point sources (Galactic stars).
The probability of each source being an extended object is indicated by the digit 0, 1, 2, 3, 4, or 5 in the ``ext\_flg'' of the catalog.
The probability increases as the digit increases (see the AllWISE documentation).
We select the sources with ext\_flg of 0,1,2 as the Galactic stars.
The selected sources form more than 99\% of all AllWISE sources. 

To all sources that satisfy the above criteria and are brighter than the $5\sigma$ sensitivity limits in the AllWISE catalog (W1$\,=16.9$ and W2$\,=16.0\,$mag), we apply Equation (7) and calculate their integrated intensities, $I_i({\rm DISL})$, assuming the averaged DIRBE beam profiles. 
These maps are described in the panels (b) and (b') of Figure 1.
We then define the ISL term as
\begin{equation}
{\it I_i({\rm ISL}) = c_i I_i({\rm DISL})},
\end{equation}
where $c_i$ is a free parameter that reflects the contributions of fainter sources than the AllWISE sensitivity limits and the effect of the photometric calibration difference between DIRBE and WISE.
Equation (10) also assumes that the ISL intensity of fainter sources (${\rm W1}>16.9$ and ${\rm W2}>16.0$ mag) has a  spatial distribution same as that of brighter sources, $I_i({\rm DISL})$.

\subsubsection{Residual Emission}

The residual emission, which includes EBL, is assumed to be independent of the region.
It is expressed as  ${\it I_i({\rm Resid})}$ and defined as
\begin{equation}
{\it I_i({\rm Resid}) = d_i},
\end{equation}
where ${\it d_i}$ is a free parameter.

\subsection{Decomposition of the Four Components}

At this stage, the model intensity, ${\it I_i({\rm Model})}$, of ${\it I_i({\rm Obs})}$ is described as
\begin{equation}
{\it I_i({\rm Model})=I_i({\rm ZL})+I_i({\rm DGL})+I_i({\rm ISL})+I_i({\rm Resid})}
\end{equation}
\begin{equation}
= {\it a_iI_i({\rm Kel})+b_iI_{{\rm 100}}+c_iI_i({\rm DISL})+d_i}.
\end{equation}

\subsubsection{Selection of the Analyzed Region}

Prior to fitting, we remove regions that might perturb the analysis.
First, the analyzed region is limited to the high Galactic latitude of $|{\it b}|>35 ^\circ$, where the optical depth at the wavelengths of interest is sufficiently small and the linear combination model (Equation (12)) is valid because there is no attenuation of $I_i({\rm Resid})$. 
In the same field, Paper I confirmed linear correlations at $1.25$ and $2.2\, \mu \rm m$.
Second, the AllWISE sources in the ecliptic longitude range of $44.7^\circ$$\sim$$54.8^\circ$ or $230.9^\circ$$\sim$$238.7^\circ$, which were observed during the early 3-band Cryo phase of the WISE mission, are reported to have missing or elevated uncertainty at the W1 band.
These regions were excluded in the analysis at  $3.5\, \mu \rm m$.
Third, Kelsall et al. (1998) pointed out that their model leaves systematic residuals in the low-ecliptic latitudes of $|{\it \beta}|<15 ^\circ$ at $4.9\, \mu \rm m$.
We therefore limit the region to $|{\it b}|>35 ^\circ$ and $|{\it \beta}|>20 ^\circ$ in the analysis at $4.9\, \mu \rm m$.

We must also conduct additional limitation.
First, the magnitudes of some bright sources in the AllWISE catalog are only given as the $2\sigma$ upper limits due to the large photometric uncertainty.
We mask the surrounding regions of such sources.
Second, the regions around the sources brighter than W1 = $4\,\rm{mag}$ are removed to reduce the photometric uncertainty.
About 30\% of the region is removed by the mask around the bright sources. 
Third, we blank out the regions around the Magellanic Clouds and probable Galactic extended sources listed in the Explanatory Supplement to the 2MASS All Sky Data Release and Extended Mission Products (Cutri et al. 2003).
Fourth, we limit our analysis to the regions of $I_{100} < 6\, \rm{MJy\,sr^{-1}}$ in which Galactic extinction is assumed to be negligible. 
Finally, outliers in the ISL intensity are excluded by applying $2\sigma$  clipping to the ISL intensities ${\it I_i(\rm{DISL})}$.

\subsubsection{Fitting Strategy}

To determine the parameters ${\it a_i}$, ${\it b_i}$, ${\it c_i}$, and ${\it d_i}$ (Equation (13)), we minimize the following ${\it \chi}^2$ function in each band:
\begin{equation}
{\it \chi_i^{\rm 2} = \sum_j\frac{[I_i({\rm Obs}) - I_i({\rm Model})]^{\rm 2}}{\sigma_i^{\rm 2}}}
\end{equation}
\begin{equation}
= {\it \sum_j\frac{[I_i({\rm Obs}) - a_iI_i({\rm Kel}) - b_iI_{{\rm 100}} - c_iI_i({\rm DISL}) - d_i]^{\rm 2}}{\sigma_i^{\rm 2}}},
\end{equation}
where ``${\it j}$'' refers to the pixels used in the fitting.
The total uncertainty, ${\it \sigma_i}$, at each pixel is calculated as follows:
\begin{equation}
{\it \sigma_i^{\rm 2} = \sigma_i({\rm Obs})^{\rm 2} + b_i^{\rm 2}\sigma_{{\rm 100}}^{\rm 2} + c_i^{\rm 2}\sigma_i({\rm DISL})^{\rm 2}},
\end{equation}
where ${\it \sigma_i({\rm Obs})}$, $\sigma_{100}$, and $\sigma_i({\rm DISL})$ are the standard deviations of the intensities of the $\epsilon = 90^\circ$ map, that of the $100\,\rm{\mu m}$ emission, and that of the ISL intensity ${\it I_i(\rm{DISL})}$, respectively.
We adopt the value $\sigma_{100} = 0.35\, \rm{MJy\,sr^{-1}}$ derived by Ienaka et al. (2013). 
The ${\it\sigma_i ({\rm DISL})}$ at each pixel is calculated in the same way as $I_i({\rm DISL})$, convolved with the DIRBE beam profile (see subsection 2.2.3):
\begin{equation}
{\it \sigma_i} ({\rm DISL})^2 = [\alpha_i 10^{-0.4 {\it m_i}}]^2 \sigma_{F^{W_i}_{{\rm 0}}}^2 \\
\end{equation}
\begin{equation}
+ [-0.4\,(\log{10})\,10^{-0.4 {\it m_i}} \alpha_i F^{W_i}_{{\rm 0}}]^2 \sigma_{{\it m_i}}^2 \\
\end{equation}
\begin{equation}
+ [F^{W_i}_{{\rm 0}} 10^{-0.4 {\it m_i}}]^2 \sigma_{\alpha_i}^2,
\end{equation}
where $\sigma_{F^{W_i}_{{\rm 0}}}$, $\sigma_{{\it m_i}}$, and $\sigma_{\alpha_i}$, respectively,  denote the uncertainties of the zero magnitude in the WISE band (4.600 and $2.560\,{\rm Jy}$ in the W1 and W2 bands, respectively (Wright et al. 2010)), that of the magnitude of each AllWISE source, and that of the conversion factor ${\alpha_i}$, set to be 1\% of ${\alpha_i}$.
Sources with no uncertainty entry in the AllWISE catalog are assigned an uncertainty, $\sigma_{{\it m_i}}$, of $0.5\,\rm{mag}$.






\section{RESULTS \& DISCUSSION}

\begin{figure*}
\begin{center}
 \includegraphics[scale=0.7]{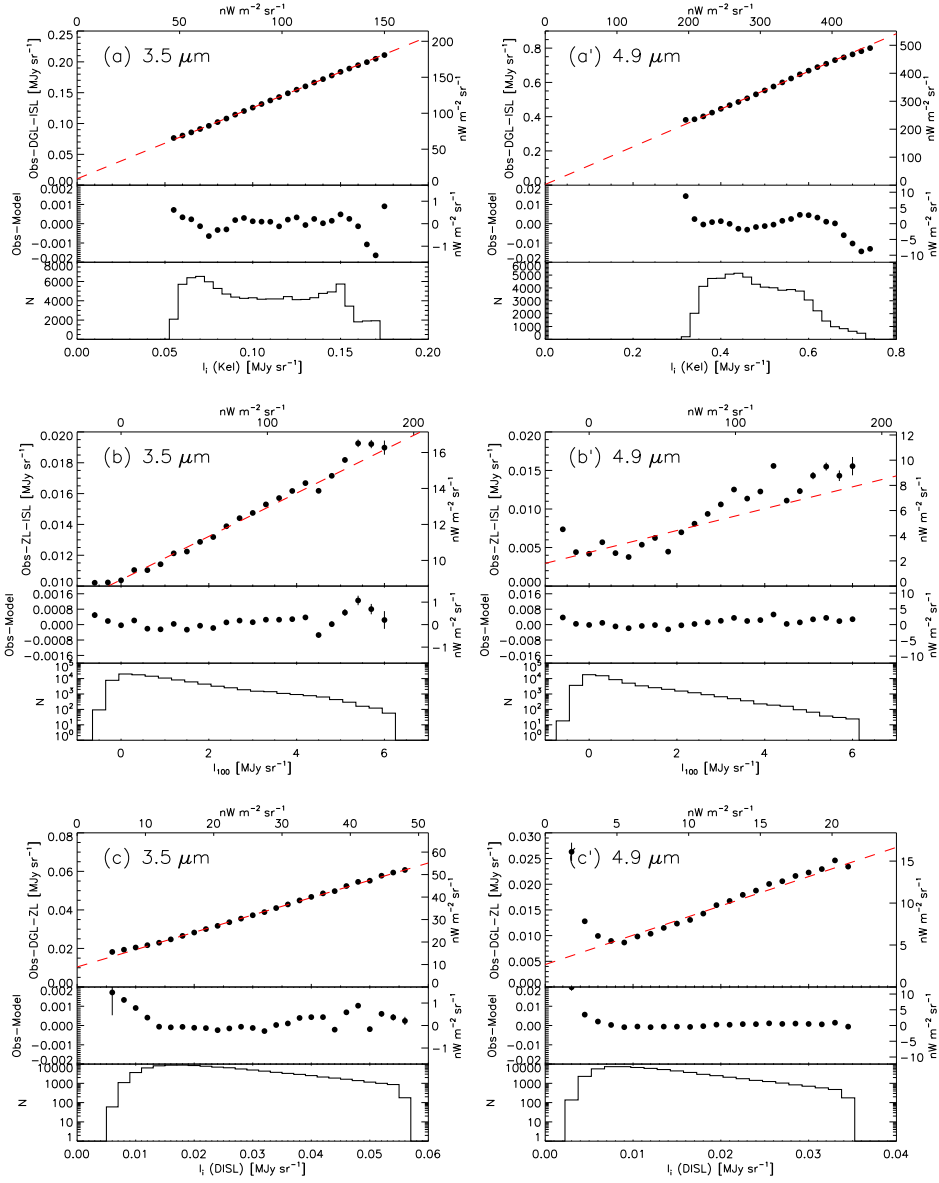} 
 \caption
 {Fitting results at $3.5\,\rm{\mu m}$ ($|b|>35 ^\circ$) and $4.9\,\rm{\mu m}$ ($|b|>35 ^\circ$ and $|\beta|>20 ^\circ$). 
  Top panels (a) and (a') plot ${\it I_i({\rm Obs}) - I_i({\rm DGL}) - I_i({\rm ISL})}$ (i.e., ${\it a_iI_i({\rm Kel}) + d_i}$) versus ${\it I_i({\rm Kel})}$; center panels (b) and (b') plot ${\it I_i({\rm Obs}) - I_i({\rm ZL}) - I_i({\rm ISL})}$ (i.e., ${\it b_iI_{{\rm 100}} + d_i}$) versus $I_{100}$; and bottom panels (c) and (c') plot ${\it I_i ({\rm Obs}) - I_i({\rm DGL}) - I_i({\rm ZL})}$ (i.e., ${\it c_iI_i({\rm DISL}) + d_i}$) versus ${\it I_i({\rm DISL})}$.
  The red lines plot the determined parameters.
  The middle and bottom parts of each panel plot the ${\it I_i({\rm Obs}) - I_i({\rm Model})}$ and the number of pixels, respectively, as functions of ${\it I_i({\rm Kel})}$ (top), $I_{100}$ (center), and ${\it I_i({\rm DISL})}$ (bottom).
  The filled circles and error bars represent the weighted means and the weighted standard errors of the sample within the arbitrary x-direction bin. 
 }
\label{fig04}
\end{center}
\end{figure*}

\subsection{Results of the Fitting}

The parameters determined by the fitting and the corresponding statistical uncertainties are listed in the ``Result'' and ``Statistical'' rows in Table 1, respectively.
Owing to the sample size of tens of thousands of pixels, the statistical uncertainty of each component is very small.
In Figure 2, we illustrate the fitting results with the determined parameters in both bands.
Each panel shows that each emission component is decomposed from the sky brightness, according to the assumed linear combination model (Equation (13)).
Filled circles and error bars in each panel, respectively, represent the weighted means and uncertainties of the points within an arbitrary x-direction bin.
In the following discussion, these weighted means are used as representative values.
In particular, Panels (b) and (b') of Figure 2 illustrate the direct linear correlation between the intensity of interstellar $100\,\rm{\mu m}$ emission and that of the diffuse near-IR light, indicating the existence of a DGL component in high Galactic latitudes.
In Table 2, we present the average of the determined intensity of each component in the analyzed region, together with its standard deviation.
In comparison with the emission components at $3.5 \,\rm{\mu m}$, the ZL accounts for more than $95\%$ of the sky brightness at $4.9 \,\rm{\mu m}$ on average.
Such a strong ZL component can make the decomposition analysis at $4.9 \,\rm{\mu m}$ more difficult and cause more dispersion of the samples from the best-fit lines in the result of the faint DGL component (panel (b') of Figure 2).

In addition to the fitting in the entire high-latitude sky, we divide the region into six Galactic longitude bins: $0 ^\circ < {\it l }< 60 ^\circ$, $60 ^\circ < {\it l} < 120 ^\circ$, $120 ^\circ < {\it l} < 180 ^\circ$, $180 ^\circ < {\it l} < 240 ^\circ$, $240 ^\circ < {\it l} < 300 ^\circ$, and $300 ^\circ < {\it l} < 360 ^\circ$.
We then conduct the $\chi^2$ minimum analysis in each field to estimate the scatter of the parameters between each region.
The determined parameters at each Galactic longitude field at $3.5$ and $4.9\,\rm{\mu m}$ are illustrated in Figures 3 and 4, respectively.
The regional difference of each parameter may be caused by the simultaneous fit of each component, which is adopted in the present analysis.
If the intensities of some components have similar spatial distributions in a region, part of the intensity of the component can be absorbed or given by that of other components.
The degree of multicollinearity is different in each region, causing regional variation in the fitting results.
This phenomenon is inevitable in the decomposition analysis of multiple components over the wide field of the sky.
As a conservative uncertainty associated with the regional variation in each parameter, we calculate the standard deviation of the determined value in the six regions and list them in the row ``Scatter'' in Table 1.
Owing to the much stronger ZL component at $4.9 \,\rm{\mu m}$, the ``Scatter'' of each parameter is larger at $4.9 \,\rm{\mu m}$ than at $3.5 \,\rm{\mu m}$.
For the parameters $a_i$, $b_i$, and $c_i$, the total uncertainties are expressed as the quadrature sum of the uncertainties of ``Statistical'' and ``Scatter'' in the row  ``Quadrature sum'' in Table 1, though the ``Scatter'' component is dominant. 

The panels (b) and (c) of Figure 3 show that all or most of the values at the six divided regions fall into one side of the all-sky value, though the values in the divided regions are naively expected to be distributed evenly around the all-sky value.
These phenomena may be attributed to the intensity difference of each emission component.
As shown in Table 2, even at $3.5 \,\rm{\mu m}$, the typical intensity of the ZL can be 10 to 100 times higher than that of the ISL and DGL.
In this situation, the small difference of the ZL intensity between each region could cause the biased fitting results in the ISL and DGL components. 

The scaling factor of the Kelsall model, the parameter $a_i$, is determined to be 10\%--15\% larger than 1.0 in both bands (Table 1), indicating that the Kelsall model underestimates the ZL brightness.
This trend has also been reported by Tsumura et al. (2013a) and Matsumoto et al. (2015) in their analysis of the {\it AKARI} and {\it IRTS} data, respectively.
As suggested by Paper I, the incompleteness of the Kelsall model can contribute to such deviations.
To determine the numerous physical parameters of ZL, Kelsall et al. (1998) used only part of the DIRBE data to avoid excessive computational times.
This approximation can lower the precision of the determined parameters in the Kelsall model, such as phase function, albedo of interplanetary dust.

The scaling factor of the ISL, the parameter $c_i$, is determined to be less than 1.0 (Table 1).
Taking into account the contributions of stars fainter than the sensitivity limits of AllWISE, the parameter $c_i$ should be more than 1.0.
However, several studies have reported a similar trend in the correlation analysis of the DIRBE data against the ISL of the 2MASS sources (Cambr\'esy et al. 2001; Levenson et al. 2007).
They reported that the parameter can be less than 1.0 by $\sim 10\%$ at $2.2\,\rm{\mu m}$.
Cambr\'esy et al. (2001) attributed this to the different methods of photometric calibration used by DIRBE and 2MASS; that is, Sirius was used as a reference for DIRBE (Hauser et al. 1998), but several faint stars were used for 2MASS (Skrutskie et al. 2006) to avoid saturation.
WISE was calibrated by several stars fainter than Vega (Wright et al. 2010), similar to the 2MASS calibration strategy.
The small value of $c_i$ can be caused because the effect of photometric calibration difference between DIRBE and WISE is more dominant than the contribution of the ISL of fainter stars .
Additionally, the very small value of $0.570\pm0.254$ at $4.9\,\rm{\mu m}$ may be contributed by the multicollinearity effect between ZL and ISL due to the intense ZL component and the very weak ISL component (Table 2).
The derived typical ISL intensities (Table 2) are consistent with the values of the Faint Source Model used in the previous analysis of the DIRBE data (see Figure 2 of Hauser et al. (1998)).
     
The determined parameters, ${\it b_i}$ (DGL) and ${\it d_i}$ (residual emission), are discussed in the following two subsections.

\begin{figure*}
\begin{center}
 \includegraphics[scale=0.6]{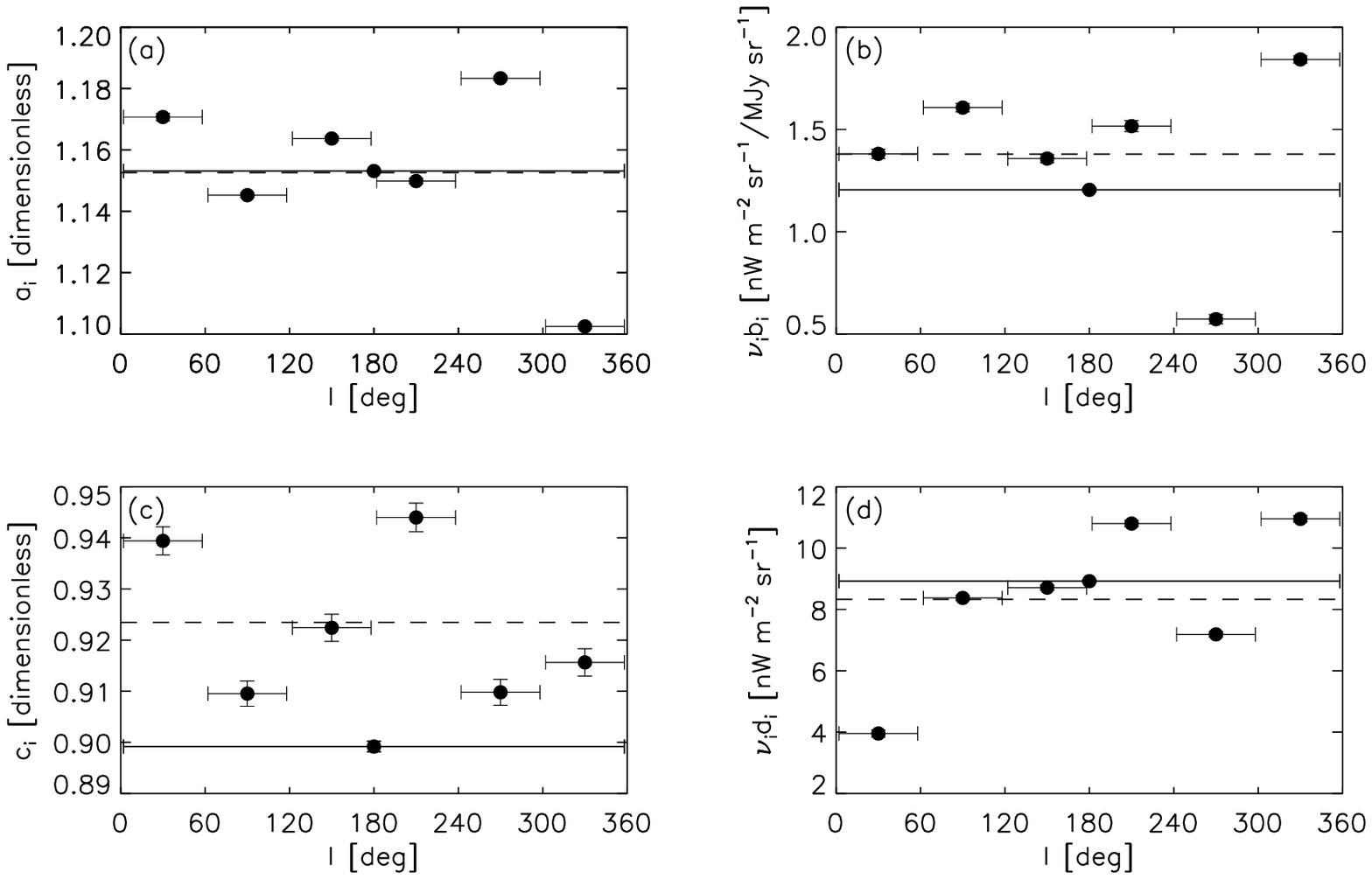} 
 \caption
 {Parameter variation among the six sampled regions as functions of the Galactic longitude at $3.5 \,\rm{\mu m}$.
 Panels (a), (b), (c), and (d) plot the determined parameters ${\it a_i}$, ${\it b_i}$, ${\it c_i}$, and ${\it d_i}$, respectively.
 Circles represent the results in each of the six regions and across the entire sky ($|{\it b}|>35 ^\circ$).
  Horizontal and vertical error bars indicate the ranges of the region and the statistical uncertainties of each parameter, respectively.
The horizontal dashed lines in each panel represent the averaged value of the determined parameters in the six sampled regions.
}
\label{fig06}
\end{center}
\end{figure*}

\begin{figure*}
\begin{center}
 \includegraphics[scale=0.6]{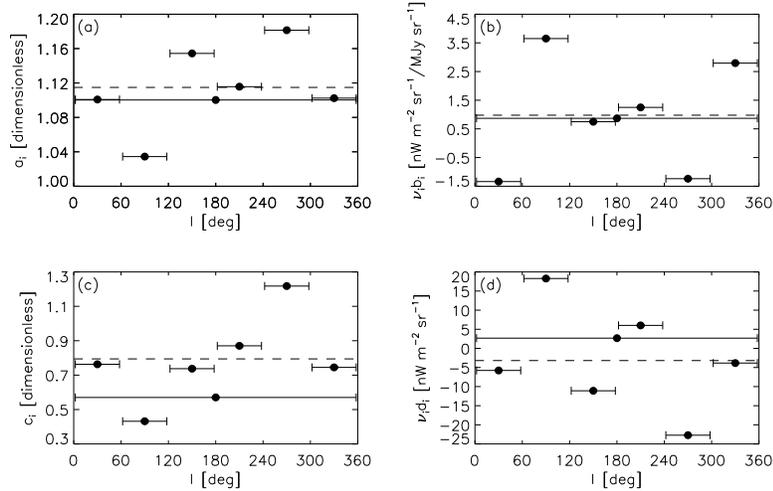} 
 \caption
 {Same as Figure 3 but at $4.9 \,\rm{\mu m}$.}
\label{fig06}
\end{center}
\end{figure*}

\begin{table*}
\begin{center}
 \renewcommand{\arraystretch}{1.0}
 \caption{Fitting results and uncertainties of each parameter}
  \label{symbols}
  \scalebox{0.9}{
  \begin{tabular}{lcccccccccccc}
  \hline
  \hline
    && \multicolumn{2}{c}{${\it a_i}$ (dimensionless)} && \multicolumn{2}{c}{${\it \nu_i b_i}$ ($\rm{nWm^{-2}sr^{-1}/MJy\,sr^{-1}}$)} && \multicolumn{2}{c}{${\it c_i}$ (dimensionless)} && \multicolumn{2}{c}{${\it \nu_i d_i}$ ($\rm{nWm^{-2}sr^{-1}}$)}\\
    \cline{3-4} \cline{6-7} \cline{9-10} \cline{12-13}
   Band ($\rm{\mu m}$) && $3.5$ & $4.9$ && $3.5$ & $4.9$ && $3.5$ & $4.9$ && $3.5 $ & $4.9$\\
 \hline
 Result   && 1.1531 & 1.1003 && 1.205 & 0.868 && 0.899 & 0.570 && 8.92 & 2.67\\
 \hline
   Statistical      && 0.0003 & 0.0002  && 0.008 & 0.010 && 0.001 & 0.002 && 0.04 & 0.05\\
   Scatter          && 0.0282 & 0.0506  && 0.433 & 2.042 && 0.015 & 0.254 && 2.59 & 14.12\\ 
   Gain             && --- & --- && --- & --- && --- & --- && 0.28 & 0.08  \\
   Galaxies         && --- & --- && --- & --- && --- & --- && 0.04 & 0.04\\
   ZL model         && --- & --- && --- & --- && --- & --- && 2.1 & 5.9 \\
   Quadrature sum   && 0.0282 & 0.0506 && 0.433 & 2.042 && 0.015 & 0.254 && 3.35 & 15.30\\
    \hline
        \end{tabular}
    }
    \end{center}
    \medskip
    
    Note. --- Symbols in the column headings are defined in subsection 2.2.\\
    The analyzed region is $|{\it b}|>35 ^\circ$ at $3.5 \,\rm{\mu m}$ and $|{\it b}|>35 ^\circ$, $|{\it \beta}|>20 ^\circ$ at $4.9 \,\rm{\mu m}$.
    
 \end{table*}

\begin{table*}
\begin{center}
 \renewcommand{\arraystretch}{1.0}
 \caption{Typical intensities of each sky emission component determined by the fitting}
  \label{symbols}
  \scalebox{0.9}{
  \begin{tabular}{lcc}
  \hline
  \hline
 
   Component ($\rm{nWm^{-2}sr^{-1}}$) & $3.5 \,\rm{\mu m} $ & $4.9 \,\rm{\mu m}$ \\
   
   \hline
   ${\it \nu_i I_i({\rm ZL}) = \nu_i a_i I_i({\rm Kel})}$ & $108.1\pm32.3$ & $334.2\pm61.6$ \\
   ${\it \nu_i I_i({\rm DGL}) = \nu_i b_i I_{{\rm 100}}}$ & $1.0\pm1.3$ &  $0.4\pm0.8$ \\ 
   ${\it \nu_i I_i({\rm ISL}) = \nu_i c_i I_i({\rm DISL})}$  & $20.0\pm8.2$ &  $5.0\pm2.3$  \\
   ${\it \nu_i I_i({\rm Resid}) = \nu_i d_i}$ & $8.9\pm3.4$ &  $2.7\pm15.3$ \\
   \hline
   ${\it \nu_i I_i({\rm Obs})}$ & $138.7\pm36.1$ &  $343.2\pm62.7$ \\
   
    \hline
    \end{tabular}
    }
    \end{center}
    \medskip
    
    Note. --- Except for ${\it I_i({\rm Resid})}$, each component is represented by its average and the standard deviation \\
    of the samples in the analyzed region --- $|{\it b}|>35 ^\circ$ at $3.5 \,\rm{\mu m}$ and $|{\it b}|>35 ^\circ$, $|{\it \beta}|>20 ^\circ$ at $4.9 \,\rm{\mu m}$.\\
    The surrounding regions of bright sources (${\rm W1}<4\,{\rm mag}$) are excluded in both bands.

 \end{table*}

\subsection{Diffuse Galactic Light in High-Galactic Latitudes}

As illustrated in panels (b) and (b') of Figure 2, we first find the DGL signal at the high Galactic latitude region at $3.5$ and $4.9\,\rm{\mu m}$ in the correlation between diffuse near-IR and $100\,\rm{\mu m}$ emission.
Figure 5 shows the current results for intensity ratios between DGL and diffuse $100\,\rm{\mu m}$ emission in the optical to the near-IR wavelengths.

\subsubsection{Model of DGL Spectra}

As indicated by the blue and green curves in Figure 5, Brandt \& Draine (2012) estimated the spectra of the starlight scattered by the plane-parallel-distributed dust grains, based on the different dust models of Zubko et al. (2004), hereafter ZDA04, and Weingertner \& Draine (2001), hereafter WD01, respectively, with the de-reddening correction of the original ISRF estimation derived by Mathis et al. (1983), hereafter MMP83, and the stellar population synthesis model of Bruzual \& Charlot (2003), hereafter BC03.
Brandt \& Draine (2012) used the BC03 model with solar metallicity and a star formation rate of $\propto \exp (-t/5{\rm Gyr})$, where $t$ denotes the timescale in units of Gyr.
As summarized in Chapter 23 of Draine (2011), both the ZDA04 and the WD01 models are composed of graphite, silicate, and PAH materials.
However, they differ in the size distributions of the grains.
The half-mass radius, $a_{0.5}$  (50\% of the total mass in grains with the radius $a>a_{0.5}$), is $0.06$ and $0.07\,\rm{\mu m}$ for the graphite and silicate grains, respectively, in ZDA04, but is $0.12\,\rm{\mu m}$ for both grains in WD01, leading to a much greater mass in $a>0.2\,\rm{\mu m}$ in WD01.
Draine (2011) pointed out that the WD01 model better reproduces the observed extinction curve from ultraviolet to the near-IR wavelengths, as derived by Fitzpatrick (1999).

In addition to the scattering component of DGL, the orange curves in Figure 5 represent the expected spectra of the near-IR emission from interstellar dust, a mixture of amorphous silicate and carbonaceous grains including PAH (Draine \& Li 2007; hereafter DL07).
As shown in Figure 12 of DL07, the increase in the PAH abundance directly causes high-intensity dust emission in the near-IR.
In Figure 5, we show the three DL07 models in which the mass fraction between PAH particles and total dust, $q_{\rm PAH}$, is different, i.e., $q_{\rm PAH} = 0.5\%$, $1.8\%$, or $4.6\%$.
To compare with the values obtained in the general interstellar fields, the scaling factor against the ISRF intensity of MMP83, $U$, is set to $U = 1$, --- corresponding to the ISRF intensity for the solar neighborhood.
In the literature, the value of $U = 1$ has been adopted in the general interstellar field throughout the sky (e.g., Dwek et al. 1997; Li \& Draine 2001; Draine \& Li 2007; Compi\`egne et al. 2011).
All model spectra in Figure 5 show the distinct PAH feature of the C-H stretching mode at $3.3\,\rm{\mu m}$.
In DL07, the intensity of the dust emission is calculated as $\lambda I_{\lambda}/N_{\rm H}\, {\rm (erg \,s^{-1} sr^{-1} H^{-1})}$, where $N_{\rm H}$ denotes hydrogen column density.
To convert $N_{\rm H}$ to the diffuse $100\,\rm{\mu m}$ intensity, we use the ratio of $100\,\rm{\mu m}$ emission to H I column density derived from the DIRBE data at high-latitude regions of $|{\it b}|>25 ^\circ$ --- 18.6 $\rm{nWm^{-2}sr^{-1}}/10^{20} {\rm cm^{-2}}$ (Arendt et al. 1998).

\subsubsection{Comparison between the DGL Model and the Observed Results}

Compared with the values of the DL07 spectra convolved with each DIRBE band (asterisks in Figure 5), the present result at $3.5\,\rm{\mu m}$ falls between the model with $q_{\rm PAH} = 4.6\%$ and that with $q_{\rm PAH} = 1.8\%$.
At $4.9\,\rm{\mu m}$, the present result prefers the model with $q_{\rm PAH} = 4.6\%$, without the large uncertainty caused by the regional variation in the decomposition analysis (see subsection 3.1).
As shown in Figure 16 of DL07, the IR emission colors obtained by the {\it Spitzer}/IRAC (Infrared Array Camera) observation toward several regions of the lower Galactic latitudes (Flagey et al. 2006) are closer to that of the DL07 model with $q_{\rm PAH} = 4.6\%$.
Compi\`egne et al. (2011) also modeled the high-latitude dust emission with the PAH parameter of $7.7\%$.
The dust emission intensity at 3--$5\,\rm{\mu m}$ is comparable between their model and the DL07 model of $q_{\rm PAH} = 4.6\%$; the difference between the two models fall within $\sim 20\%$ in that wavelength range (see Figure 6 of Compi\`egne et al. (2011)). 
Combining the present values with these results, it is probable that $q_{\rm PAH}$ is above  $\sim 2\%$ in the high-latitude region.

In the previous analysis of DIRBE data, Arendt et al. (1998) derived the intensity ratios of DGL to $100\,\rm{\mu m}$ emission only in low-latitude regions of $|{\it b}|<30 ^\circ$ (filled black circles in Figure 5).
At $3.5\,\rm{\mu m}$, their low-latitude result is comparable to the present high-latitude value. 
This indicates that the PAH abundance do not greatly change throughout the general interstellar field in the Milky Way, except in the region of the Galactic plane.
At $4.9\,\rm{\mu m}$, the result of Arendt et al. (1998) is two times higher than that of the model spectra; Li \& Draine (2001) thus suggested an additional opacity for the ultra-small grain component to explain this excess.
However, such components may not be required by our result at $4.9\,\rm{\mu m}$, without the uncertainty of regional variation.

The intensities of the DL07 models with $q_{\rm PAH} = 4.6\%$ are a few times lower than that obtained with {\it AKARI} (Tsumura et al. 2013b; solid black curve in Figure 5) in the continuum spectra.
The low-latitude result of {\it AKARI} ($5^\circ<|{\it b}|<15^\circ$) may cause such a difference.
It is suspicious that $U = 1$ can be adopted in the low-latitude region because of its higher ISRF.
Unfortunately, Tsumura et al. (2013b) did not find the DGL feature in the high-latitude region, probably due to the low signal-to-noise ratio of the observation, same as in the reanalysis of the {\it IRTS} data (Matsumoto et al. 2015).

To maintain consistency with the results at optical wavelengths (Matsuoka et al. 2011; Ienaka et al. 2013), Paper I and Arai et al. (2015) have both suggested a bluer DGL spectrum at 1--$2\,\rm{\mu m}$, fitted to the twice-scaled ZDA04 model spectra (blue curves in Figure 5).
However, there is no firm reason to allow the scaling of the model to fit the observed results.
In addition, the size distribution of the dust grains differs between the DL07 and ZDA04 models; DL07 adopts the WD01 model, which better reproduces the observed extinction curve (Fitzpatrick 1999).
Therefore, it may be premature to conclude that the scattering components of DGL have a bluer spectrum with a larger contribution of smaller dust grains.
Figure 5 shows that the original model spectra based on the WD01 model (green curves) can be fitted to the observed results of Paper I and Arai et al. (2015) without violating the present results at $3.5$ and $4.9\,\rm{\mu m}$.
In this case, the observed results of Matsuoka et al. (2011) are larger than the model spectra by a factor of 2.
To explain this discrepancy, the albedo of the dust grains in the optical wavelengths should be larger than the DGL model of Brandt \& Draine (2012) without a change in the albedo in the near-IR.
By solving the simple radiative transfer of light through the dust slab, Ienaka et al. (2013) expressed the parameters $b_i$ as functions of albedo and optical depth (see Appendix of Ienaka et al. 2013).
According to the solution of their equation, only an $\sim 20\%$ increase in the visual albedo can enhance the parameter $b_i$ by a factor of 2.
This small change in the optical albedo in the model of Brandt \& Draine (2012) may explain the excess values of Matsuoka et al. (2011).

For a further study of interstellar dust using DGL, both precise spectroscopic observations of DGL with high signal-to-noise ratios and DGL models which simultaneously reproduce scattering and thermal emission components will be needed.

\begin{figure*}
\begin{center}
 \includegraphics[scale=0.8]{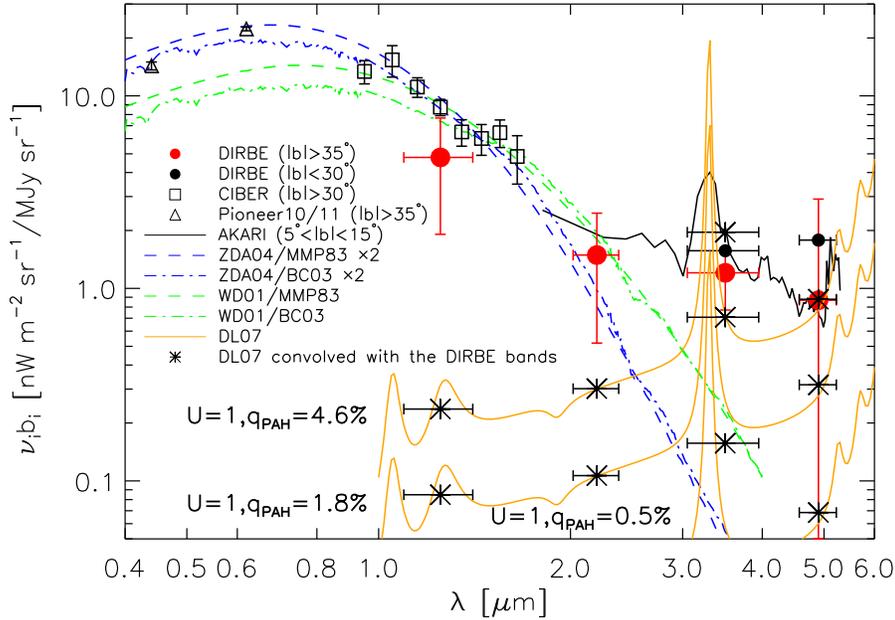} 
  \caption
 {Intensity ratios of DGL to $100\,\rm{\mu m}$ emission ${\it \nu_i b_i}$ as functions of wavelength.
 The results of the DIRBE reanalysis at high Galactic latitude are plotted as filled red circles (present study and Paper I), with the horizontal error bars denoting the bandwidth of DIRBE.
The other results are obtained with {\it Pioneer 10/11} (Matsuoka et al. 2011; open triangles), CIBER (Arai et al. 2015; open squares), and {\it AKARI} (Tsumura et al. 2013b; solid black curve), and a previous analysis of the DIRBE data (Arendt et al. 1998; filled black circles).
The blue and green curves are the spectra of the scattering component of DGL, estimated by Brandt \& Draine (2012) using the WD01/BC03 (green dash-dotted curve), WD01/MMP83 (green dashed curve), ZDA04/BC03 (blue dash-dotted curve), and ZDA04/MMP83 (blue dashed curve) models. 
To fit the observed results at the optical wavelengths (Matsuoka et al. 2011), ZDA04-based models are scaled by two times.
Solid orange curves represent the spectra of the interstellar dust emission estimated by DL07 with the ISRF and PAH parameters of $U = 1$, $q_{\rm PAH} = 4.6\%$ (the upper curve), $U = 1$, $q_{\rm PAH} = 1.8\%$ (the middle curve), and $U = 1$, $q_{\rm PAH} = 0.5\%$ (the lower curve).
Asterisks indicate the values of the DL07 model spectra convolved with each DIRBE band.
}
\label{fig06}
\end{center}
\end{figure*}

\subsection{Residual Emission Component}

\subsubsection{Uncertainty Estimation of the Residual Emission}

For the residual emission components $d_i$, we estimate the additional uncertainties associated with the absolute gain of DIRBE, faint galaxies, and the ZL model.

Hauser et al. (1998) reported uncertainties of $3.1\%$ and $3.0\%$ in the absolute gain of DIRBE at $3.5$ and $4.9 \,\rm{\mu m}$, respectively.
The values of these uncertainties correspond to percentages of the derived parameter $d_i$, and appear in the row ``Gain'' in Table 1.

The AllWISE catalog may contain unresolved faint galaxies, which should be added to the uncertainty budget of the parameter $d_i$.
Levenson et al. (2007) estimated the contribution of such galaxies at $3.5 \,\rm{\mu m}$ of $0.04\, \rm{nWm^{-2}sr^{-1}}$, which corresponds to galaxies of $K_s < 14.3\,\rm{mag}$ in the 2MASS PSC.
Assuming this value is of the same order as the AllWISE sources, we adopt this as the uncertainty of faint galaxies at $3.5 \,\rm{\mu m}$.
Generally, the spectrum of a galaxy does not drastically change between $3.5$ and $4.9 \,\rm{\mu m}$.
We then set the same value at $4.9 \,\rm{\mu m}$.
These uncertainties are listed in the ``Galaxies'' row in Table 1.

As described in Kelsall et al. (1998), the uncertainty of the Kelsall model is estimated as the intensity difference between the two ZL models at the north Galactic pole (NGP) where the difference is reported to be the largest.
This uncertainty is $2.1$ and $5.9\,\rm{nWm^{-2}sr^{-1}}$ at $3.5$ and $4.9 \,\rm{\mu m}$, respectively, listed in the row ``ZL model'' in Table 1.

The quadrature sum of the uncertainties of the parameter $d_i$ is presented in the row ``Quadrature sum'' in Table 1.
For the parameter $d_i$, the uncertainties associated with the regional variation in the parameter and the Kelsall model dominate over the other uncertainties.

\begin{figure*}
\begin{center}
 \includegraphics[scale=0.6]{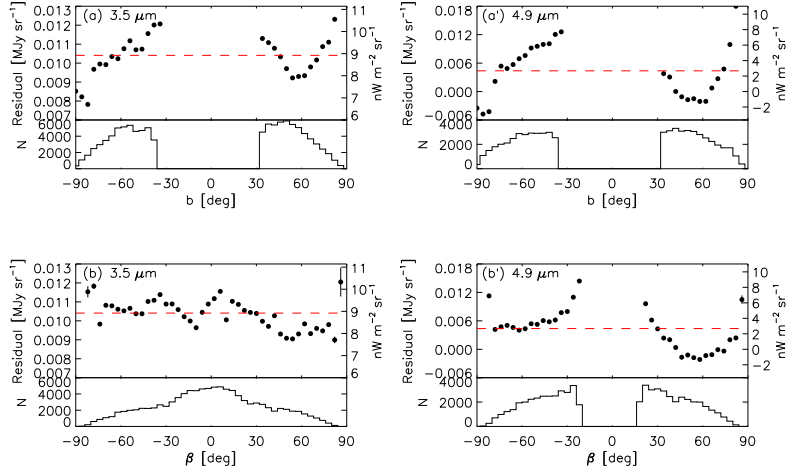} 
 \caption
 {Dependence of the residual emission on the Galactic and ecliptic latitudes, determined by the fitting in the region of $|{\it b}|>35 ^\circ$ at $3.5\,\rm{\mu m}$ and $|{\it b}|>35 ^\circ$, $|{\it \beta}|>20 ^\circ$ at $4.9\,\rm{\mu m}$.
The upper part of each panel plots ${\it I_i({\rm Obs}) - I_i({\rm ZL}) - I_i({\rm ISL}) - I_i({\rm DGL})}$ as a function of Galactic latitude ${\it b}$ (top panels) and ecliptic latitude ${\it \beta}$ (bottom panels), represented by the weighted means of the points within arbitrary x-direction bins (filled circles).
Horizontal red dashed lines represent the determined parameters $d_i$.
The lower part of each panel is a histogram of the number of pixels at each 
 ${\it b}$ or ${\it \beta}$.}
\label{fig04}
\end{center}
\end{figure*}

\begin{figure*}
\begin{center}
 \includegraphics[scale=0.6]{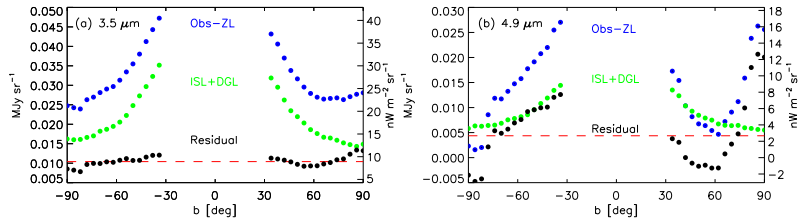} 
 \caption
 {Galactic latitude dependence of the intensities of the residual emissions ${\it I_i({\rm Obs}) - I_i({\rm ZL}) - I_i({\rm ISL}) - I_i({\rm DGL})}$ (black circles), ${\it I_i({\rm Obs}) - I_i({\rm ZL})}$ (blue circles), and ${\it I_i({\rm ISL}) + I_i({\rm DGL})}$ (green circles) at $3.5\,\rm{\mu m}$ (panel (a)) and $4.9\,\rm{\mu m}$ (panel (b)).
Filled circles denote the same quantities as those in Figure 6.
Horizontal red dashed lines are the determined parameter $d_i$.
}
\label{fig04}
\end{center}
\end{figure*}

\subsubsection{Isotropy of the Residual Emission}

To test the isotropy of the residual emission, Figure 6 illustrates ${\it I_i({\rm Obs}) - I_i({\rm ZL}) - I_i({\rm ISL}) - I_i({\rm DGL})}$ with the parameters $d_i$ being functions of the Galactic latitude ${\it b}$ (panels (a) and (a')) and the ecliptic latitude ${\it \beta}$ (panels (b) and (b')) in both bands.

Other than in the region where ${\it b} \gtrsim 60^\circ$, the residual emission shows similar trends at $3.5$ and $4.9 \,\rm{\mu m}$ as functions of Galactic latitude; it increases toward the low-Galactic latitudes (panels (a) and (a')).
Paper I found the same trend in the analysis of the residual emission at $1.25$ and $2.2 \,\rm{\mu m}$ with the 2MASS data and suggested two explanations for this trend.
One was the simply modeled ISL term, ${\it I_i({\rm ISL}) = c_iI_i({\rm 2MASS})}$, where $I_i({\rm 2MASS})$ denotes the integrated intensity of the 2MASS sources below the detection limit. 
Such a trend could be caused if the spatial distribution of intensity as a function of Galactic latitude differs between $I_i({\rm 2MASS})$ and the ISL of stars fainter than the detection limits of 2MASS.
Another was the contribution of the faint stars that were not detected in the 2MASS PSC due to the masking effect of the nearby bright sources. 
Such faint stars increase as the number density of the bright sources increases toward the lower-Galactic latitudes.
These explanations can be applied to the present study, which evaluates ISL using the AllWISE sources in the same way as the 2MASS PSC in Paper I.

In the region of ${\it b} \gtrsim 60^\circ$, the residual emission shows the inverse behavior: it increases toward NGP.
To investigate this phenomenon, we plot ${\it I_i({\rm Obs}) - I_i({\rm ZL})}$ and ${\it I_i({\rm ISL}) + I_i({\rm DGL})}$ with the residual emission in Figure 7.
Reasonably, the modeled Galactic components ${\it I_i({\rm ISL}) + I_i({\rm DGL})}$ increase toward the low-Galactic latitudes.
In contrast, ${\it I_i({\rm Obs}) - I_i({\rm ZL})}$ shows the same feature as the residual emission  in the ${\it b} \gtrsim 60^\circ$ region, indicating that the trend is caused by $I_i({\rm ZL})$.
The fact that such a feature is larger at $4.9 \,\rm{\mu m}$ than at $3.5 \,\rm{\mu m}$ is then reasonable because of the stronger ZL component at $4.9 \,\rm{\mu m}$.
However, the reason why such a feature is seen only at ${\it b} \gtrsim 60^\circ$ is unclear.
For one thing, Kelsall et al. (1998) pointed out that the intensity differences between the different ZL models are largest in the NGP region.
Such difficulty in the modeling of ZL near the NGP region may be related to the trend at ${\it b} \gtrsim 60^\circ$.

As a function of ecliptic latitude, the residual emission is relatively constant at $3.5 \,\rm{\mu m}$, but is systematically larger toward the lower-ecliptic latitudes at $4.9 \,\rm{\mu m}$ (panels (b) and (b') of Figure 6).
This indicates the difficulty of modeling of ZL at $4.9 \,\rm{\mu m}$, where the ZL intensity is much stronger than that at $3.5 \,\rm{\mu m}$.
In addition to the high intensity of the ZL at $4.9 \,\rm{\mu m}$, the incompleteness of the ZL model in the band may contribute to the large scatter of the other components --- DGL, ISL, and the residual emission (see Figure 4).
In this situation, there seems to be room for improvement on the ZL model, though it is generally difficult.

Though the residual emissions show some amount of scatter as functions of ${\it b}$ and ${\it \beta}$, they are within the typical scatter derived by the regional variation in the parameters $d_i$ (see Table 1 and subsection 3.1), because such regional variation naturally includes the dependence of ${\it b}$ and ${\it \beta}$. 
Such scatter as functions of ${\it b}$ and ${\it \beta}$ is also comparable to the estimated systematic uncertainty of the Kelsall model (Table 1).
For the isotropy measurement of the residual emissions, we therefore use the ``Scatter'' value (Table 1) as the conservative values.
In our analysis, the deviation of the residual emission from the isotropy is then less than $\sim 30\%$ of the residual intensity at $3.5 \,\rm{\mu m}$.
This deviation from the isotropy is larger than that of $\lesssim 10\%$ at $1.25$ and $2.2 \,\rm{\mu m}$ (Paper I), partly because the residual emission intensity at $3.5 \,\rm{\mu m}$ is more than two times smaller than those at $1.25$ and $2.2 \,\rm{\mu m}$.
We do not discuss the isotropy of the residual emission at $4.9 \,\rm{\mu m}$ due to the very large scatter.

\subsubsection{Comparison with Other Studies}

In Figure 8, we compare the resultant residual emissions of Paper I and the present study with those of previous studies.
As illustrated in Figure 5, the present DGL result at $3.5 \,\rm{\mu m}$ in the high Galactic latitudes is comparable to the results obtained at low-Galactic latitudes (Arendt et al. 1998) within the uncertainty.
This leads to the same level of residual emissions as in the previous studies adopting the DGL result derived by Arendt et al. (1998) at $3.5 \,\rm{\mu m}$ (Gorjian et al. 2000).
At $4.9 \,\rm{\mu m}$, the residual emission is not significantly detected due to large uncertainty associated with the ZL subtraction, same as the previous studies (Hauser et al. 1998; Tsumura et al. 2013c).

The intensity of the residual emission changes according to the different ZL models; using the Kelsall model or Wright's (1998) model, several studies have reported large residual emissions at $1.25$ and $2.2 \,\rm{\mu m}$ (Gorjian et al. 2000; Cambr\'esy et al. 2001; Levenson et al. 2007; Tsumura et al. 2013c; Matsumoto et al. 2015; Paper I), which are $2$--$5$ times the observed or modeled IGL intensity (e.g., Madau \& Pozzetti 2000; Totani et al. 2001; Stecker et al. 2006; Mazin \& Raue 2007; Franceschini et al. 2008; Finke et al. 2010; Dom\'inguez et al. 2011).
Because such large residual emissions also exceed most of the $\gamma$-ray constraints of EBL (e.g., Aharonian et al. 2006; Abramowski et al. 2013), it is difficult to regard them as being entirely of extragalactic origin.
In addition, Paper I suggested that IGL, together with the contributions of all of the exotic extragalactic sources estimated so far, including Pop-III stars (Inoue et al. 2013; Fernandez et al. 2013), IHL (Zemcov et al. 2014), direct collapse black holes (Yue et al. 2013), and dark stars (Maurer et al. 2012), cannot account for the excess residual emission, especially at $1.25 \,\rm{\mu m}$.
Therefore, it is probable that part of the excess emission comes from the local universe.

Compared with the residual emissions at $1.25$ and $2.2 \,\rm{\mu m}$, the present result at  $3.5 \,\rm{\mu m}$ is relatively small, approaching the IGL level and $\gamma$-ray constraints.
Combined with the results of Paper I, the spectrum of the near-IR residual isotropic emission has the Rayleigh-Jeans spectrum, consistent with the previous studies.
In contrast, a large gap exists between the two results in the optical wavelengths: Matsuoka et al. (2011) and Bernstein (2007).
Because the results of Matsuoka et al. (2011) were obtained from the {\it Pioneer 10/11} observation beyond 3AU from the Earth, where the ZL contribution is negligible, the residual emission may be regarded as the optical EBL intensity.
If this is true, the intensity gap between Matsuoka et al. (2011) and Bernstein (2007) can be attributed to the unmodeled ZL contribution.

The present decomposition analysis, which deals with the integrated light of each emission component cannot identify where the residual emission comes from.
As mentioned above, part of the residual emission can arise from the local universe, such as the Milky Way or solar system due to the contradiction against the $\gamma$-ray constraints.
In their interpretation of the derived isotropic residual emissions at $140$ and $240 \,\rm{\mu m}$, Dwek et al. (1998) suggested that the Galactic component cannot produce the residual intensity because an unreasonably large gas and dust mass is needed.
From this point of view, the ``isotropic DGL component'', the counterpart of the isotropic far-IR emission from ISM, is unlikely to contribute to the residual emission in the optical to near-IR wavelengths.
Recently, Lehner et al. (2015) reported the presence of the massive circumgalactic medium around the Andromeda galaxy.
However, it is very unclear that such component also exists around the Milky Way and creates the isotropic emission.

Another possible origin of the residual emission may be within the solar system, including ZL.
As reported by Dwek et al. (2005), the spectrum of near-IR excess residual emission is similar to the ZL spectrum.
In addition, some isotropic components associated with ZL can be added to the Kelsall model because absolute measurement of ZL is impossible from the orbit of the Earth (Hauser et al. 1998).
However, such isotropic ZL components have not been observationally confirmed and it is suspicious that they would have the same spectrum as the currently measured ZL. 
To reveal whether the residual emission components include the ZL contribution, the EXo-Zodiacal Infrared Telescope (EXZIT), one of the science instruments of the Solar Power Sail spacecraft planned for launch in 2020s, will be useful; this mission is planned to observe the universe in the near-IR wavelengths beyond the interplanetary dust region (Matsuura et al. 2014).
EXZIT will reveal the three-dimensional structure of interplanetary dust and the absolute EBL intensity without the uncertainty of the ZL model.

\begin{figure*}
\begin{center}
 \includegraphics[scale=0.8]{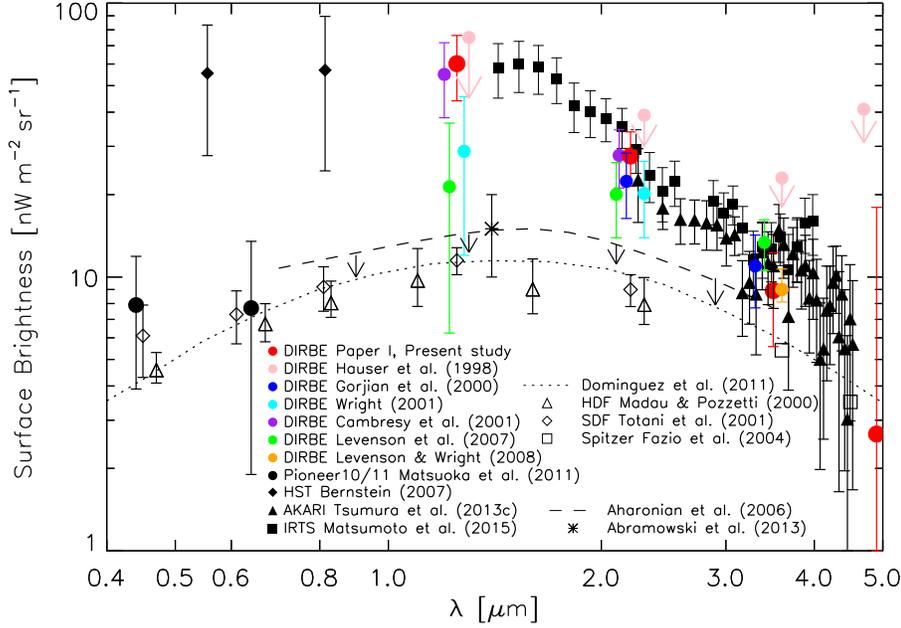} 
 \caption
 {Current measurements of the residual emission from the optical to the near-IR. 
  Filled red circles are the results of the present study and Paper I.
 The other colored symbols are the results of previous {\it COBE}/DIRBE data analyses conducted by Hauser et al. (1998) (pink down-arrows), Cambr\'esy et al. (2001) (purple circles), Levenson et al. (2007) (green circles), Gorjian et al. (2000) (blue circles), Wright (2001) (cyan circles), and Levenson \& Wright (2008) (orange circle). 
 The other results were obtained from {\it HST}/WFPC2 (Bernstein 2007; filled black diamonds), {\it IRTS} (Matsumoto et al. 2015; filled black squares), {\it AKARI} (Tsumura et al. 2013c; filled black triangles), and {\it Pioneer 10/11} (Matsuoka et al. 2011; filled black circles).
  Open diamonds, triangles, and squares are the IGL intensity obtained from the {\it Subaru} Deep Field (Totani et al. 2001), {\it Hubble} Deep Field (Madau \& Pozzetti 2000), and {\it Spitzer}/IRAC (Fazio et al. 2004), respectively. 
  The dotted curve indicates the estimated spectrum of IGL (Dom\'inguez et al. 2011).
  The EBL upper limits estimated by the $\gamma$-ray observation is indicated by the dashed curve (Aharonian et al. 2006).
   The recent constraint by H.E.S.S. at $1.4\,\rm{\mu m}$ is indicated by the asterisk (Abramowski et al. 2013).
 For clarity, some results are shifted a little from their exact wavelengths.
}
\label{fig07}
\end{center}
\end{figure*}

\section{SUMMARY}

We reanalyze the diffuse sky brightness at 3.5 and $4.9\,\rm{\mu m}$ using the {\it COBE}/DIRBE data in the high Galactic latitude region, in which DGL evaluation has been controversial for EBL measurements due to the low optical depth to enhance the DGL intensity.
Since Paper I succeeded in finding the DGL component at 1.25 and $2.2\,\rm{\mu m}$ with the precise ISL evaluation by  2MASS, we adopt a similar method at 3.5 and $4.9\,\rm{\mu m}$ using the AllWISE source catalog instead of 2MASS.

Taking account of the DIRBE beam shapes for the AllWISE sources below the sensitivity limit, we create the ISL intensity maps at 3.5 and $4.9\,\rm{\mu m}$, same as Paper I.
We then assume that the sky brightness is expressed as a linear combination of ZL, DGL, ISL, and residual emission including the EBL components and decompose the four components by a $\chi^2$ minimum analysis.
As a result, we first find the direct linear correlation between diffuse near-IR light and interstellar $100\,\rm{\mu m}$ emission in the high-latitude region of $|{\it b}| > 35^\circ$, indicating the extraction of the DGL component.
The high-latitude DGL result at $3.5\,\rm{\mu m}$ is revealed to be comparable to the low-latitude value derived from the previous DIRBE analysis (Arendt et al. 1998).

Compared with the model of the DGL spectra that assumes a size distribution of the dust grains composed of amorphous silicate and graphite including PAH particles (DL07), the present results at $3.5$ and $4.9\,\rm{\mu m}$ constrained the mass ratio of PAHs to the total dust grains to be above $\sim 2\%$, which is consistent with the results of {\it Spitzer}/IRAC.

For the residual emission, we derive the weak result of $8.9\pm3.4\,\rm{nWm^{-2}sr^{-1}}$ at $3.5\,\rm{\mu m}$, consistent with other studies.
Compared with the results of the residual emissions at 1--$2\,\rm{\mu m}$, this intensity at $3.5\,\rm{\mu m}$ approaches the $\gamma$-ray constrains on EBL and the IGL level, showing that the residual emission intensity exhibits the Rayleigh-Jeans spectrum in the near-IR.
In our analysis, the deviation of the residual emission from the isotropy was less than $30\%$ at $3.5\,\rm{\mu m}$.
At $4.9\,\rm{\mu m}$, the uncertainty of the residual emission is large, but the result in this band is consistent with the upper limits derived by the previous studies.

The ISL maps created by the AllWISE sources at $|{\it b}| > 20^\circ$ are available in the online version of this journal.




\acknowledgments

We wish to thank T.~D. Brandt and K. Tsumura for providing their data. 
We are grateful to the referee, Varoujan Gorjian, for a number of useful comments that improved the paper. 
K.S. is supported by Grant-in-Aid for Japan Society for the Promotion of Science (JSPS) Fellows. 

This publication uses the {\it COBE} datasets developed by the National Aeronautics and Space Administration (NASA) Goddard Space Flight Center under the guidance of the {\it COBE} Science Working Group.
This publication also makes use of data products from WISE, which is a joint project between the University of California, Los Angeles, and the Jet Propulsion Laboratory/California Institute of Technology.
WISE is funded by NASA.






\appendix




\clearpage



\clearpage









\clearpage

\end{document}